\documentclass[%
 reprint,
 amsmath,amssymb,
 aps,
]{revtex4-1}

\usepackage{graphicx}% Include figure files
\usepackage{dcolumn}% Align table columns on decimal point
\usepackage{bm}% bold math
\usepackage[T1]{fontenc}

\usepackage{mathptmx}      % use Times fonts if available on your TeX system
\usepackage{flushend}
\usepackage{esvect}

\def\be{\begin{equation}}
\def\ee{\end{equation}}
\def\bea{\begin{aligned}}
\def\eea{\end{aligned}}
\def\ba{\begin{eqnarray}}
\def\ea{\end{eqnarray}}

\def\yzero{\smash{\hbox{$y\kern-4pt\raise1pt\hbox{${}^\circ$}$}}}

\def\-{\hphantom{-}}

\def\s2{\frac{1}{\sqrt2}}

\def\IF{\relax{\rm I\kern-.18em F}}
\def\II{\relax{\rm I\kern-.18em I}}
\def\IP{\relax{\rm I\kern-.18em P}}
\def\IC{\relax\hbox{\kern.25em$\inbar\kern-.3em{\rm C}$}}
\def\IR{\relax{\rm I\kern-.18em R}}

\def\Dsl{\,\raise.15ex\hbox{/}\mkern-13.5mu D} %this one can be subscripted
\def\IZ{Z\kern-.4em  Z}

\begin{document}

\preprint{APS/123-QED}

\title{Classification of  M2-brane 2-torus bundles, U-duality invariance and type II  gauged supergravities.}% Force line breaks with \\
%\thanks{A footnote to the article title}%

%\title{T-duality invariance, M2-brane bundles and type II classification of gauged supergravities\thanksref{t1}}
%\author{Maria Pilar Garcia del Moral\thanksref{e1,addr1} \and J.M. Pena\thanksref{e2,addr2} \and Alvaro Restuccia\thanksref{e3,addr1,addr4}}

\author{Maria Pilar Garcia del Moral}
 \email{maria.garciadelmoral@uantof.cl}
 \affiliation{Departamento de F\'isica, Universidad de Antofagasta, Aptdo 02800, Chile.}%Lines break automatically or can be forced with \\

\author{J.M. Pena}%
 \email{jpena@fisica.ciens.ucv.ve}
\affiliation{Departamento de F\'isica,  Facultad de Ciencias,
 Universidad Central de Venezuela, A.P. 47270, Caracas 1041-A, Venezuela.}
 \author{Alvaro Restuccia}
 \email{alvaro.restuccia@uantof.cl}
 \affiliation{ Departamento de F\'isica, Universidad de Antofagasta, Aptdo 02800, Chile.}%Lines break automatically or can be forced with \\

\begin{abstract}

In this paper we obtain the complete classification of the inequivalent classes of M2-brane symplectic torus bundles with monodromy in $SL(2,Z)$ and their precise U-duality relations among them. There are eight inequivalent classes of bundles whose monodromy groups, at low energies, are in correspondence with the gauging groups of the eight type II gauged supergravities in nine dimensions. Four of those have been previously found and they correspond to the 'type IIB side'. In this paper we provide the explicit realization of the remaining four classes associated to the 'type IIA side'.  The precise  M2-brane U-duality relations between the eight inequivalent classes of bundles have allowed to identified the remaining four ones. We conjecture that the classes of  gaugings -classifying the eight types of II gauged supergravity in nine dimensions- are determined by the inequivalent coinvariant classes associated to the base and the fiber of the supermembrane bundles and their duals.


%\begin{description}
%\item[Usage]
%Secondary publications and information retrieval purposes.
%\item[PACS numbers]
%May be entered using the \verb+\pacs{#1}+ command.
%\item[Structure]
%You may use the \texttt{description} environment to structure your abstract;
%use the optional argument of the \verb+\item+ command to give the category of %each item.
%\end{description}
\end{abstract}

%\pacs{Valid PACS appear here}% PACS, the Physics and Astronomy
                             % Classification Scheme.
%\keywords{Suggested keywords}%Use showkeys class option if keyword
                              %display desired
\maketitle

%

%\institute{Departamento de F\'isica, Universidad de Antofagasta, Aptdo 02800, Chile\label{addr1}
%          \and
%         Departamento de F\'isica,  Facultad de Ciencias, Universidad Central de Venezuela, A.P. 47270, Caracas 1041-A, Venezuela\label{addr2}
%           \and
%Departamento de F\'\i sica, Universidad Sim\'on Bol\'\i var, Apartado 89000, Caracas 1080-A, Venezuela\label{addr4}
%}

\date{Received: date / Accepted: date}
% The correct dates will be entered by the editor

\maketitle

\section{Introduction}
M-theory is a theory candidate for unification of all the interactions in Nature that contains Supermembrane theory -also called M2-brane theory- as one of its building blocks. Any quantum consistent definition of M-theory will require of its  understanding.  Supermembranes are $2+1$ dimensional objects embedded in $11D$ space-time that act as sources for $D=11$ Supergravity and as a consequence of it there is a deep relation between both theories: on one hand Supermembrane theory is conjectured to contain a  unique massless groundstate associated to the $11D$ supergravity multiplet. Several works have been developed in support of this claim, for a recent new approach see \cite{gmbr5} and references therein. 
On the other, Supermembrane theories  are expected to be described by supergravities at  'low energies'.  In that respect, the M-theory origin of maximal supergravities in any dimension $d\le 11$ \cite{julia1, julia2}, is well known to correspond to the $11D$ supermembrane compactified on a trivial $T^{11-d}$ torus \cite{bst}.  

There are other type of Supergravity theories like gauged/ massive supergravities. Gauged supergravities can be obtained from String/Supergravity theories by a number of ways: by compactifying on manifolds with nontrivial holonomy \cite{nicolai, gates},  flux compactifications \cite{samtleben}, gauging procedures like the Embedding tensor \cite{trigiante, melgarejo-ortin}, Scherk-Schwarz reductions (SS) \cite{berg} where the reduced fields keep a non-trivial phase dependence on the  internal spatial coordinates \cite{tp, patrick, bergshoeff}, or as effective descriptions of String/M-theory formulated on nontrivial torus bundles \cite{gates-zwiebach1, gates-zwiebach2, bedwn}. In the torus bundle formulation  the  internal dependence of SS fields becomes  associated to a twist given by a monodromy ${\mathcal{M}}_{\mathtt{sugra}}$ of the bundles  \cite{hull-massive, llp2, lavrinenko1}.

In distinction, the determination of the M-theory action origin of the gauged/massive supergravity deformations has become much more elusive.  The M-theory uplift of SS nine dimensional reduction was conjectured to be related to torus bundles with monodromy in $SL(2,Z)$ \cite{hull-massive}.  The action must be an invariant functional formulated in terms of the local sections of this bundle. In \cite{gmpr} the authors showed  that the supermembrane formulated on 2-torus bundles with monodromy in $SL(2,Z)$  is the origin of type II gauged supergravities  in nine dimensions. Type II  supergravities in nine dimensions consist of a unique maximal supergravity and eight  gauged deformations \cite{tp, patrick,bergshoeff} four of them coming from the type IIA sector  and  the remaining four from the type IIB one. At effective level, their  10D origin are the type IIB supergravity, the maximal type IIA and the two type IIA massive deformations: Romans \cite{Romans} and Howe-Lambert-West (HLW) \cite{hlw}. In this paper we extend the analysis of the M2-brane 'T-duality' transformation done in \cite{gmpr} where the four inequivalent M2-brane 2-torus bundles associated to the 'type IIB' sector were found an the other four were inferred from the 'T-duality' invariance of the Mass operator. We provide an explicit construction of the four inequivalent classes of M2-brane bundles associated to the 'type IIA side' in nine dimensions.

Another topic that has received a lot of attention from the community are the T-dual/U-dual invariant theories in String and M-theory in order to have a better insight of the non perturbative structure of these theories. U-duality is a nonperturbative transformation defined in 11D that uplifts and unifies the perturbative $T$-duality and nonperturbative $S$-duality present in 10D. For the case of M-theory compactified on a 2-torus U-duality group is conjectured to be $SL(2,Z)\times Z_2$. M-theory as a unification theory should contain them as  symmetries, and its effective description should be invariant.  In the context of Effective Field Theories, they have been constructed in terms of modified supergravity actions enriched with terms of stringy origin in which T-duality is manifest.  These attempts have focused mainly in two different approaches: Double Field Theory \cite{Hull-Zwiebach, tseytlin, duff,  bmt, duff-lu, daniel} and Generalized Geometry \cite{Itching, stw, reids} focusing either on its stringy action \cite{stw, 24hull-tfolds} or on the M-theory realization \cite{grana1, Berman-Perry, bmp}.  Recently it has been even possible to incorporate the $\alpha^{'}$ corrections into the analysis \cite{carmen} as well as winding contributions from the very beginning \cite{grana}. See \cite{aldazabal1, aldazabal2, aldazabal3, carmen1} for a review.

Global aspects of the T-duality are relevant and  should also be considered in order to achieve U-dual invariant actions at the level of M-theory. Some of them were studied in \cite{lozano, 24hull-tfolds}. %
In this paper we want to  exhaustively characterize the M2-branes on 2-torus bundle (with and without monodromy) which is the simplest nontrivial example of  U-dual invariant theories -at the level of Mass operators- connected with type II  maximal \textit{and} gauged supergravities.  We deep on the characterization of the M2-brane bundle description, extending the results found in \cite{gmpr}. The supermembrane bundle class is specified through the coinvariants of the fiber and the base. We emphasize the role played by the coinvariant of the base manifold in the bundle structure when the U-duality action  is performed. We will denote 'S-duality' or 'T-duality' in quotes in the text when we want to stress the particular action of 11D U-duality that connects with its counterpart (S, T-duality) when reduced to 10D. As we will see in section II, 'S-duality' is a symmetry of the Mass operator of the supermembrane, hence to verify the U-duality properties we will focus on the 'T-duality' action. 'T-duality' maps M2-brane torus bundles into M2-brane  torus bundles preserving the invariance of  the Hamiltonian and mass operator but interchanging the cohomological charges of the base manifold with the homological ones of the fiber at the same time that  the monodromy is mapped into a dual monodromy that belongs to the same conjugation class. 
We analyze the case where the U-duality group corresponds to $SL(2,Z)\times Z_2$ and  the U-dual invariant orbit of charges is classified by the coinvariant class of the M2-brane bundle. This orbit is completely filled with the charges associated to M2-branes. Consequently under U-duality the M2-brane 2-torus bundles are mapped into M2-brane 2-torus bundles and U-duality does not include charges associated to branes of different dimensionality, in distinction with the case when more compactified dimensions are considered \cite{stelle, cederwall}.
U-duality interchanges those invariants, however the duality map among the coinvariants does not correspond to an equivalence relation.  U-duality does not act on generic grounds linearly. The U-duality action on the global structure of the bundle is the relevant one to explain the difference observed at effective level between the two sectors of type II gauged supergravity in 9D from an M-theory point of view. On the supergravity side the gauging groups of the type IIA and type IIB  do not coincide so U-duality do not preserve them either \cite{tp, patrick}. This fact is totally natural at the level of the effective field theory since the global $SL(2,R)$ symmetry is not realized on the type IIA side at perturbative level. We discuss their possible relation with the M2-brane torus bundles in section IV, and in the discussion section.

The paper is structured as follows: In section II, we review the formulation of the Supermembrane theory on torus bundles with monodromy in $SL(2,Z)$ and we show that the mass operator is 'S-dual' invariant. In section II.A, we extend  the analysis previously done for the M2-brane 2-torus bundle by studying the role played by the coinvariant $C_B$ associated to the base manifold.  In section III, we describe in detail the 'T-duality' transformation for the supermembrane bundle, understanding by it the U-duality part associated to T-duality in 10D when the M2-brane theory is double dimensionally reduced. We specify its local and global action on the 2-torus bundles according to its coinvariant classification. In section IV, we obtain explicitly, by analyzing the action of 'T-duality' on the coinvariant structures, the four inequivalent classes of bundles of M2-brane bundles. We conjecture that they are associated at low energies with the gaugings in the type IIA supergravity sector in nine dimensions. In section V, we present the discussion on the role of the U-duality part associated when double dimensionally reduced to T-duality to explain the differences in the bundle structure and in the deformations allowed at supergravity level from the supermembrane point of view, and finally we present our conclusions. In the Appendix A, we describe some of the technical properties of the coinvariants of bundles with monodromy and in the Appendix B, we deduce the most general supermembrane 'T-duality' transformation.
%%%%%%%%%%%%%%%%%%%%%%%%%%%%%%%%%%
%
%
%

\section{Supermembrane theory on a symplectic torus bundle}  
In this section we will review the supermembrane theory compactified on an $M_9\times T^2$  formulated globally on symplectic torus bundles. These bundles are classified according to two inequivalent topological sectors: they can be principal (i.e. with trivial monodromy) or non trivial with a monodromy group contained in $SL(2,Z)$. The supermembrane theory in the Light Cone Gauge (LCG) has a residual gauge symmetry, the symplectomorphisms on the base manifold which in two dimensions are equivalent to the area preserving diffeomorphisms (APD). When the theory is formulated globally the group of symplectomorphisms corresponds to the structure group of the torus bundle. It is well known that supermembrane theory on a torus when doubled dimensionally reduced corresponds to the type IIB superstring compactified on a circle \cite{schwarz, sl2z}. The supermembrane compactified on a trivial torus is associated  at low energies to a type II maximal supergravity in nine dimensions \cite{schwarz}. When the compactification is non trivial, but associated to a central charge condition the theory is described at low energies by the type IIB gauged supergravities as shown in \cite{gmpr}. 

Let us review the construction of the symplectic M2-brane torus bundle in this first part of the section. The hamiltonian of a supermembrane theory with central charges formulated in the Light Cone Gauge (LCG) on a target space $M_9\times T^2$, the supermembrane subject to a central charge condition, is the following one \cite{mor1, dwhn, mr, gmr, mrt, bgmr, bgmr3}:
{\small\small

\begin{equation}
\begin{aligned}
H=&\int_{\Sigma}{T_{\mathtt{{M2}}}^{2/3}}\sqrt{\rho}\left[\frac{1}{2}(\frac{P_{n}}{\sqrt{\rho}})^2 +\frac{1}{2}\frac{P\overline{P}}{\sqrt{\rho}}+\frac{T_{\mathtt{{M2}}}^{2}}{4}\{X^{m},X^{n}\}^2\right]+\\
+&\int_{\Sigma}{T_{\mathtt{{M2}}}^{2/3}}\sqrt{\rho}\left[\frac{T_{\mathtt{{M2}}}^{2}}{2}(\mathcal{D}X^{n})(\mathcal{\overline{D}}X^{n})\right]+\int_{\Sigma}{T_{\mathtt{{M2}}}^{2/3}}\sqrt{\rho}\left[+\frac{T_{\mathtt{{M2}}}^{2}}{4}(\mathcal{F}\overline{\mathcal{F}})\right]+\\
+&({\mathtt{n}}^2 \, Area_{T^2}^2)+\\
+&\int_{\Sigma}{T_{\mathtt{{M2}}}^{2/3}}\sqrt{\rho}\left[-\overline{\Psi}\Gamma_{-}\Gamma_{n}\{X^{n},\Psi\}
-\frac{1}{2}\overline{\Psi}\Gamma_{-}\overline{\Gamma}\mathcal{D}\Psi-\frac{1}{2}\overline{\Psi}\Gamma_{-}\Gamma\overline{\mathcal{D}}\Psi\right]\\
+&\int_{\Sigma}\sqrt{\rho} \, {L} \left[\frac{1}{2}\overline{\mathcal{D}}(\frac{P}{\sqrt{\rho}})+\frac{1}{2}\mathcal{D}(\frac{\overline{P}}{\sqrt{\rho}})+\{X^{n},\frac{P_{n}}{\sqrt{\rho}}\}-\{\overline{\Psi}\Gamma_{-},\Psi\}\right],
\end{aligned}
\end{equation}}

\noindent where $L$ is a Lagrange multiplier and ${T_{\mathtt{{M2}}}}$ is the 11D tension of the supermembrane, $\rho$ is the determinant of  non-flat two torus $\Sigma$ that corresponds to the spatial part of the worldvolume metric. The symplectic bracket is defined as $\{\textsf{A},\textsf{B}\}=\omega^{ab}\partial_a \textsf{A}\partial_b \textsf{B}$ whose symplectic 2-form is $\omega=\omega^{ab}d\sigma_a\otimes d\sigma_b$ with $\omega^{ab}=\frac{i\epsilon^{ab}}{2\sqrt{\rho}}$ with $a,b=z,\overline{z}$  the local complex coordinates and  respectively its complex conjugate $\overline{z}$, defined on the base manifold $\Sigma$. %
$X^{n}$ are the embedding maps $\Sigma\to M_9$ where ${n}=3,\dots,9$ and $X=X^1+iX^2$ are the embedding ones from $\Sigma\to T^2$. They are scalars parametrizing the transverse coordinates of the supermembrane in the target space. $P_{n}$ are densities and they are the canonical momenta associated to the $X^{n}$, and respectively $P$ that of the field $X$. $\Psi$ are scalars on the worldvolume but an $SO(7)$ spinor on the target space, $\Gamma_n$ are seven Gamma matrices and $\Gamma= \Gamma_1+i\Gamma_2$, denoting by $\overline{\Gamma}$ its complex conjugate.  The 2-torus $T^2$ of the target space  is characterized by the moduli ${\mathtt{R}}$ the radius, and $\tau$ the complex Teichm\"uller parameter. The winding numbers  are $l_s,m_s$ with $r,s=1,2$ associated to the wrapping of the supermembrane on the $T^2$. They define a matrix $\mathbb{W}=\begin{pmatrix} l_{1}& l_{2}\\
               m_{1} & m_{2}
\end{pmatrix}$. When the wrapping is irreducible its determinant $\mathtt{n}$ is different from zero \cite{mrt} and the theory has discrete spectrum \cite{bgmr} in distinction with the wrapped supermembrane without this topological condition. This sector defines a topological condition associated to the existence of worldvolume monopoles that algebraically imply the existence of a non-vanishing central charge in the supersymmetric algebra. For this reason this sector was called {\textit supermembrane with central charges}.
On this sector there is a symplectic curvature defined on the base manifold is $\mathcal{F}=D\overline{A}-\overline{D}A+\{A,\overline{A}\}$, with $A$ a connection under the infinitesimal symplectomorphism transformation $\delta_{\epsilon}A=\mathcal{D}\epsilon$. See \cite{mor1, gmmpr} for a detailed analysis. The symplectic covariant derivative is defined as $\mathcal{D}\bullet=D\bullet+\{A,\bullet\}$ , with $D\bullet=e_r^a\partial_a\bullet$ a rotated covariant derivative \cite{sculpting, gmpr} defined in terms of a zwei-bein $e_r^a$ as
\begin{equation}
\label{theta}
e_r^a:=-2\pi {\mathtt{R}}(l_r+m_r\tau)\Theta_{sr}\omega^{ba}\partial_b\widehat{ X^s},
\end{equation}
with $r,s=1,2$. $d\widehat{X}$ are the harmonic one-form basis defined on $\Sigma$. The Hamiltonian is invariant the residual symmetry under Area Preserving Diffeomorphisms (APD) connected and not connected to the identity.

The hamiltonian is subject to the APD group residual constraints (connected to the identity $\phi_1$, but also to the large APD  $\phi_2$)
\begin{eqnarray}
\phi_1&:&d(\frac{1}{2}(Pd\overline{X}+\overline{P}dX)+P_mdX^m-\overline{\Psi}\Gamma_{-}\Psi)=0 \, ;\\
\phi_2&:&\oint_{\mathcal{C}_r}[\frac{1}{2}(Pd\overline{X}+\overline{P}dX)+P_mdX^m-\overline{\Psi}\Gamma_{-}d\Psi]=0 \, ,
\end{eqnarray}
where $\mathcal{C}_s$ is the canonical 1-homology basis on $T^2$.

The theory is invariant under two different $SL(2,Z)$ discrete symmetries: The first one is associated to the invariance under the change of the basis of the harmonic one forms defined on the Riemann worldvolume \cite{sl2z} and the windings
\begin{equation}
d\widehat{X}\to Sd\widehat{X},\quad \mathbb{W}\to S^{-1}\mathbb{W} \, ,
\end{equation}
with $S\in SL(2,Z)$. This dependence is encoded in the matrix $\Theta\in SL(2,Z)$ \cite{sculpting} in (\ref{theta}).

The second one is associated to an invariance of the mass operator involving $SL(2,Z)$ symmetry related to the target 2-torus $T^2$, so it is invariant under S-duality transformations,
\begin{eqnarray}
\label{transformations}
\tau &\to& \frac{a\tau +b}{c\tau +d} \,,\nonumber \\
{\mathtt{R}} &\to&  {\mathtt{R}}|c \tau +d|\,, \nonumber \\
A &\to& A e^{i\varphi_{\tau}}\, ,  \\
\mathbb{W} &\to& \begin{pmatrix} a & -b\\
                        -c & d
 \end{pmatrix} \mathbb{W}\,, \nonumber \\
{\mathbf{Q}} &\to& \begin{pmatrix} a & b \\
                        c & d
 \end{pmatrix}{\mathbf{Q}}\nonumber\, ,
\end{eqnarray}
where $\left( \begin{array}{cc} a & b \\ c & d \end{array}\right) \in SL(2,{Z})$ and $c\tau+d= |c\tau+d|e^{-i\varphi_{\tau}}$ and ${\mathbf{Q}}=\left(\begin{array}{cc} p \\ q \end{array}\right)$ is the KK charge of the supermembrane propagating on the target 2-torus considered. The homological charges of the target torus $H_1(T^2)$ are interpreted as the quantized KK charges of the compactified supermembrane \cite{gmpr}.\newline
Now we formulate the previous embedding description in terms of a symplectic torus bundle with monodromy in $SL(2,Z)$. This global formulation is going to make manifest some topological invariants that carry physical information. The total space $E$ is defined in terms of a fiber $F=M_9\times T^2$ and $\Sigma$ as the base manifold. The structure group $\mathtt{G}$ is the symplectomorphisms leaving invariant the canonical symplectic structure in $T^2$. The action of $\mathtt{G}$ on $F$ produces a $\pi_0(\mathtt{G})$-action on the homology and cohomology of $F$. The \textit{monodromy} ${\mathcal{M}}_G$ is defined as
\begin{equation}
{{\mathcal{M}}_G}:\pi_1(\Sigma) \to \pi_0(\mathtt{G})\,,
\end{equation}
with
\begin{equation}
\mathtt{G}=Symp(T^2) \quad \textrm{and} \quad \pi_0(\mathtt{G})=SL(2,Z)\, .
\end{equation}
Consequently ${{\mathcal{M}}_G}=\begin{pmatrix}a&b\\c&d\end{pmatrix}^{\gamma}\in SL(2,Z)$  and it acts on the homology basis of the $T^2$ target torus -where $\gamma={\gamma}_1+{\gamma}_2$ with $({\gamma}_1,{\gamma}_2)$ are the integers characterizing the element of the homotopic group $\pi_1(\Sigma)$. The symplectic connection defined on the base manifold transforms with the monodromy, $dA\to dA e^{i\varphi_{{\mathcal{M}}_G}}$ where ${\varphi}_{{{\mathcal{M}}_G}}$ is a discrete monodromy phase given by $\varphi_{{{\mathcal{M}}_G}}=\frac{c\tau+d}{\vert c\tau+d\vert}$ for a given modulus $\tau$. The inequivalent classes of symplectic torus bundles over $\Sigma$ are classified by the elements of the second cohomology group, {\small $H^2(\Sigma,Z^2_{{\mathcal{M}_G}})$ } or equivalently by their coinvariants. See \cite{gmmpr, khan, gmpr} for more details.
%%%%%%%%
%\newline
The global symmetries of the theory become restricted by the monodromy.

\subsection{The role of fiber $C_F$ and base $C_B$ coinvariants on the bundle structure}
The supermembrane symplectic torus bundles are characterized by two types of coinvariants relevant for the characterization of the supermembrane bundle. The class of coinvariants associated to the fiber $C_F$ and the coinvariants associated to the base $C_B$.
The torus bundles with a given monodromy ${\mathcal{M}}_G$ are classified according to the elements of the twisted second cohomology group $H^2 (\Sigma, Z_{{\mathcal{M}}}^2)$ of the base manifold $\Sigma$. Its coefficients are defined on the module generated by the monodromy representation acting on the homology of the target torus \cite{khan}. There is a bijective relation with the elements of the coinvariant group $C_F=\{C_{\mathbf{a}}\}, {\mathtt{a}}=1,\dots,j $ associated with a particular monodromy group ${\mathcal{M}}_G$, see appendix A, for more properties of the coinvariant classes.  A coinvariant class in the KK sector is given by
\be
C_{F}=\{{\mathbf{Q}}+({\mathcal{M}}_g-\mathbb{I})\widehat{{\mathbf{Q}}}\} \, ,
\label{coinvariantclasskk}
\ee
for any $g\in {\mathcal{M}}_G$, and $\widehat{{\mathbf{Q}}}$ is any arbitrary element of the KK sector. That is, two elements belong to the same class if they differ in an element $(g-1)\widehat{\mathbf{Q}}$ for some $g\in {\mathcal{M}}_G $ and $\widehat{\mathbf{Q}}$.

Associated to the monodromy subgroup ${\mathcal{M}}_G$  there is an induced action on the cohomology of the base manifold \cite{gmpr4}, which corresponds to the monodromy group of the winding sector  ${\mathcal{M}}_G^*$. Since ${\mathcal{M}}_G^{*}=\Omega {\mathcal{M}}_G \Omega^{-1}$ with $\Omega= \begin{pmatrix}-1& 0\\ 0 & 1 \end{pmatrix}$, it lies in the same conjugacy class of ${\mathcal{M}}_G$. ${\mathcal{M}}_G^*$ acts on the fields which define the Hamiltonian, that is, on sections of the torus bundle through  a matrix $\Theta=(V^{-1}{\mathcal{M}}_G^*V)^T$  that appears in the symplectic covariant derivatives $\mathcal{D}_r$ of the Hamiltonian $H$. $V$ is associated to the monodromy subgroup ${\mathcal{M}}_G\subset SL(2,Z)$, see \cite{gmpr} for further details.
%The matrix $\Xi$ plays a role in the supermembrane bundle characterization analogous to the embedding tensor in gauged supergravities.(frase %repetida)
Induced by them there are other two possible invariants $({\mathcal{M}}_G^*,C_B)$ characterizing the symplectic bundles. The two monodromies lie in the same conjugation class, however they are not the same and consequently their respective coinvariants are not equivalent.
%%%%%%%%%%%%%%%%%%%%%%%%%%%%%%%%%%%%%%%%%%%%%%

A coinvariant class in the winding sector is given by
\begin{equation}
C_B=\{{\mathbf{W}}+({\mathcal{M}}_g^*-\mathbb{I})\widehat{{\mathbf{W}}}\} \,,
\end{equation}
with ${\mathcal{M}}_g^*\in {\mathcal{M}}_G^*$, the monodromy group acting in the winding sector specified by ${\mathbf{W}}=\left(\begin{array}{cc} l \\ m \end{array}\right)\in H^1(\Sigma)$. Due to the $SL(2,Z)$ symmetry on the equivalence class of the basis of homology on the base manifold (a torus), it is possible to reduce the problem to work indistinctly with the winding matrix $\mathbb{W}$ or with ${\mathbf{W}}$ defined by the first row of the matrix $\mathbb{W}$. See appendix A.  Then for M2-brane or symplectic torus bundles one needs to specify both types of coinvariants for a given monodromy ${\mathcal{M}}_G$,  those associated to the fiber $C_F$ and those associated to the base manifold $C_B$.
Given a symplectic torus bundle the Hamiltonian of the theory and the mass operator are defined on the coinvariant classes.

The dependence of the bundle on the winding charges $\mathbf{W}$ and KK charges ${\mathbf{Q}}$ is defined in terms of the function $\textsf{F}$:
\begin{equation}
\begin{aligned}
\textsf{F}:\mathbf{W} \to \quad\mathbf{W} \,, \qquad
  {\mathbf{Q}} \to \quad {\mathbf{Q}} \, ,
\end{aligned}
\end{equation}
which depends on the classes $C_B$ and $C_F$. Since these are invariant under the action of the monodromy, the same occur for the Hamiltonian or mass operator. We thus have,
\begin{equation}
\textsf{F}(C_B)= \textsf{F}(\mathbf{W}), \quad \textsf{F}(C_F)= \textsf{F}({\mathbf{Q}}).
\end{equation}
In particular if we consider the action only on the orbits, instead of the coinvariant classes, $\textsf{F}$ may be defined in terms of a matrix $\Theta$ acting on the matrix $\mathbb{W}:\Theta \mathbb{W}$ in a way that under the monodromy $\mathbb{W}\to {\mathcal{M}}^{*} \mathbb{W}$ and $\Theta \to \Theta ({\mathcal{M}}^{*})^{-1}$ as we have defined the coinvariant derivative.
A way to obtain a function $\textsf{F}$ with such property is to consider a linear function ${\textsf{F}}_L$ defined on the orbits of the elements of the coinvariant class:
\begin{equation}
\begin{aligned}
&{\textsf{F}}_L(C_B)={\textsf{F}}_L(\mathbf{W}+({\mathcal{M}}^*-\mathbb{I})\widehat{\mathbf{W}})={\textsf{F}}_L(\mathbf{W}).
\end{aligned}
\end{equation}
Each coinvariant class is invariant under the action of any $g\in {\mathcal{M}}_G$. So the coinvariant class may be considered itself as a class of orbits under the action of ${\mathcal{M}}_G$. Given a symplectic torus bundle the Hamiltonian of the theory is defined for any orbit of the coinvariant class.
%%%%%%%%%%%%%%%%%%%%%%%%%%%%%%%%%%%%%%%%%%%%%%%%%%%%%%%%%%%%%%%%%%%%%%%%%%%%
%
%
%
\section{'T-duality' for supermembrane theory torus bundles}
Disclaimer: In this section we analyze the action of U-duality over the M2-brane torus bundle that in 10D corresponds to T-duality  when the M2-brane is doubled dimensionally reduced. In other words we focus on the part of the U-duality action that exchange winding charges and Kaluza-Klein charges, acts on the moduli at the same time that transforms the structure of the torus bundle.  In order to emphasize these aspects we also called it  under the name of 'T-duality' even when we are working  at the level of 11D. 

The 'T-duality' transformation acts on the Hamiltonian $H$ and the mass operator ${\bf M}^2$ and it has also a action on the topological invariants of the M2-brane torus bundle describing the theory. In what follows, we are going to describe the 'T-duality' action at global level, that is, specifying its action on the supermembrane symplectic torus bundle structure and secondly we will describe its action on the mass operator of the supermembranes.

\subsection{'T-duality' action on the coinvariants}
The 'T-duality' transformation globally transforms a bundle into a dual one, by interchanging the cohomological charges of the torus base manifold into the homological charges defined on the torus fiber with dual moduli. 'T-duality' also interchanges the coinvariant class of the base and the fiber in the dual Torus-bundle:
\begin{equation}
(C_F,C_B)=({\widetilde{C}}_B, {\widetilde{C}}_F),
\end{equation}
where $\widetilde{C}$ denotes the dual coinvariant class. It may occur, however that the transformation becomes non linear. At low energies this fact will be reflected in the change of the gauging group associated to the corresponding dual supergravity. In order to delve in this classification we are going to characterize the action of 'T-duality' over the different classes of M2-brane bundles with monodromies trivial and non trivial:
%%%%%%%%%%%%%%%%%%%%%%%%%%%%%%%%%%%%%%%%%%%%%%%%%%%%%%%%%%%%%%%%%%%%%%%%%%%%%%%%%%%%%%%%%%%%%%%%%
%
%
\begin{itemize}
\item {\em Trivial monodromy}:
We first consider the case in which ${\mathcal{M}}_G=\mathbb{I}$, i.e. when the monodromy group is trivial. In this case the coinvariant classes, which classify the inequivalent torus bundles, have only one element $\mathbf{Q}$ in the KK sector and one element $\mathbf{W}$ in the winding sector. The 'T-dual' transformation is defined in terms of $\mathcal{T}\in SL(2,Z)$ with equal diagonal terms, satisfying
\be
\label{dualcharge}
\widetilde{Q}=\mathbf{W}= \mathcal{T}\mathbf{Q}, \, \qquad \widetilde{W}=\mathbf{Q}= \mathcal{T}^{-1} \mathbf{W} \, ,
\ee
where $\mathcal{T}=\left( \begin{array}{cc} \alpha & \beta \\ \gamma & \alpha \end{array}\right) \in SL(2,Z)$ is defined as in the appendix C.
Given $\mathbf{Q}$ and an associated  winding matrix $\mathbb{W}$, there always exists a winding matrix on the equivalence class of $\mathbb{W}$ defined  by the action from the right by $\mathbb{S}\in SL(2,Z)$ such that
\be
\mathbb{W} \mathbb{S}=\mathcal{T} \mathbb{Q} \, ,
\ee
$\mathbb{Q}$ is the matrix whose first row is $\mathbf{Q}$ and it also has determinant $\mathtt{n}$. Given $\mathbf{Q}$ there always exists $\mathbb{Q}$, though it is not unique. The most general one is obtained by multiplying from the right by a parabolic matrix with integer coefficients and equal diagonal elements:
\be
\mathbb{Q}\mathbb{K} \, , \quad \mathbb{K}=\begin{pmatrix}1& k\\ 0& 1\end{pmatrix}.
\ee
%
%
%%%%%%%%%%%%%%%%%%%%%%%%%%%%%%%%%%%%%%%%%%%%%%
%
%
The matrix $\mathbb{K}$ can always be absorbed into $\mathbb{S}$. The symplectic torus bundles are classified in this case, i.e. ${\mathcal{M}}_G=\mathbb{I}$, by two integers, the elements $Z\otimes Z$. The symplectic torus bundles are in one to one correspondence with the $U(1)\times U(1)$ principle bundle over the base manifold. Since the monodromy is trivial, the structure group may reduce to the group of symplectomorphism homotopic to the identity. The dual transformation is then completed by the transformation of the moduli as given in (\ref{tt1}).

\item {\em Non trivial monodromy}: We now consider the case where the monodromy group $\mathcal{M}_G$ is non trivial. It is an abelian subgroup of $SL(2,Z)$.  The 'T-dual' transformation maps as before coinvariant classes on the KK sector onto coinvariant class in the winding sector. In order to define the 'T-duality' map we take any element $\mathbf{Q}$ of $C_F$ and any element $\mathbf{W}$ of $C_B$ and map them as in (\ref{dualcharge}). The global 'T-duality' map is given by the following transformation,
\be
\bea 
&\mathbf{Q}\stackrel{\mathcal{T}}{\rightarrow} \mathbf{W} \, , \\
&{\mathcal{M}}_G\stackrel{\Omega}{\rightarrow} \Omega \,  {\mathcal{M}}_G \,\Omega^{-1}  \, , \\ &\widehat{\mathbf{Q}}\stackrel{f}{\rightarrow}\widehat{\mathbf{W}}\, ,  \\
& C_F\stackrel{\mathbb{T}}{\rightarrow} C_B \,,
\eea\ee
where $f$ is a general linear map from the $\widehat{\mathbf{Q}}$ sector onto the $\widehat{\mathbf{W}}$ sector. We denote by $\mathbb{T}$ the 'T-duality' action on the coinvariants that can act linearly $\mathcal{T}$ or not. In particular the map $f$ can be defined as:
\begin{equation}
\widehat{\mathbf{W}}=\mathcal{T} \widehat{\mathbf{Q}}.
\end{equation}
However, in general it is not necessary to relate $f$ to $\mathcal{T}$.

Suppose now that instead of mapping $\mathbf{Q}$ to $\mathbf{W}$ we map it to another member of the coinvariant class to which $\mathbf{W}$ belongs:
\be
\label{w}
\mathbf{Q}\stackrel{\mathcal{T}}{\rightarrow} \mathbf{W}+({\mathcal{M}}_G^*-\mathbb{I})\widehat{\mathbf{W}}_1  \, ,
\ee
then,
\be
\begin{aligned}
\label{member}
&C_F=\{\mathbf{Q}+({\mathcal{M}}_G-\mathbb{I})\widehat{\mathbf{Q}}\}
 \stackrel{\mathbb{T}}{\rightarrow}\{\mathbf{W}+(\mathcal{M}_G^*-\mathbb{I})(\widehat{\mathbf{W}}+\widehat{\mathbf{W}}_1)\}=C_B.
\end{aligned}
\ee
That is, the new map is only a translation on the $\widehat{\mathbf{W}}$ sector, the  map $f$ includes a translation by $\widehat{\mathbf{W}}_1$. Equation (\ref{member}) shows that changing the map for $\mathbf{Q}\to \mathbf{W}$ to $\mathbf{Q}\to (\mathbf{W}+({\mathcal{M}}_G^*-\mathbb{I})\widehat{\mathbf{W}})$ is equivalent to leave the map $\mathbf{Q}\to \mathbf{W}$ and change $f$ by a translation. The translation by $\widehat{\mathbf{W}}_1$ is irrelevant from the point of view of the coinvariant class since $\widehat{\mathbf{W}}+\widehat{\mathbf{W}}_1$ is a general element of the winding sector. Hence the map between the coinvariant classes is only determined by $\mathcal{T}$ which is constructed from one element of each class $\mathbf{Q}$ and $\mathbf{W}$ respectively. The generator ${\mathcal{M}}_G$ can be parabolic, elliptic or hyperbolic.
%We will analyze these cases separately in the section 5.
\end{itemize}
\subsection{'T-duality' action on the mass operator}
The duality transformation on the symplectic torus bundle has an action on the charges but also on the geometrical moduli. We define dimensionless variables $\mathcal{Z}$, where $\mathcal{Z}=(T_{\mathtt{M2}} \mathtt{A} Y)^{1/3}$ with $\mathtt{A} = (2\pi \mathtt{R})^2 Im\tau$, the area of the target torus and $Y=\frac{\mathtt{R} Im\tau}{\vert q\tau -p\vert}$ a variable proportional to the $\mathtt{R}$ radius of the complex torus.  $Y$ is invariant under the monodromy group if we consider $\mathbf{Q}$ the components of ${\textsf{F}}(\mathbf{Q})$.  The 'T-duality' transformation is given by:
\begin{equation}
\begin{aligned}
\label{tt1}
&\textrm{The moduli}: \quad \mathcal{Z}\widetilde{\mathcal{Z}}=1,\quad\widetilde{\tau}=\frac{\alpha \tau+\beta}{\gamma\tau +\alpha}. \\
&\textrm{The charges}: \quad \widetilde{\mathbf{Q}}=\mathcal{T} \mathbf{Q},\quad \widetilde{\mathbf{W}}=\mathcal{T}^{-1}\mathbf{W} \, ,
\end{aligned}
\end{equation}
with $\alpha, \beta, \gamma$, the integer entries of the $\mathcal{T}$ matrix in (\ref{dualcharge}). The charges $\mathbf{Q}, \mathbf{W}$ transform depending on the type of bundle considered, i.e. with trivial or non trivial monodromy. We notice that $\mathcal{Z},Y$ and their duals are invariant on an orbit generated by ${\mathcal{M}}_G$ contained in the respective coinvariant class, provided that $\tau$ and $\mathbf{Q}$ transform as in (\ref{transformations}). Moreover, they are independent of the coinvariant class when we define $Y$ in terms of the components of ${\textsf{F}}(\mathbf{Q})$ instead of $\mathbf{Q}$ and leave $\tau$ as a invariant moduli under monodromy.  The symmetry of the Hamiltonian related to the basis of harmonic one-forms of the Riemann worldvolume \cite{sl2z} allows to define the class of orbits associated to the winding matrices $[\mathbf{W}]$. Following (\ref{dualcharge}) there always exists $\mathcal{T}$ such that
\be
\mathcal{T}\mathbf{Q}=[\mathbf{W}],\quad \mathcal{T}^{-1}\mathbf{W}=[\mathbf{Q}]  \, ,
\ee
such that 'T-duality' maps classes into classes
\be
[\mathbf{W}]\to [\widetilde{\mathbf{Q}}]=[\mathbf{W}],\quad [\mathbf{W}]\to [\widetilde{\mathbf{W}}]=[\mathbf{Q}] \, .
\ee

Let us recall \cite{gmpr,gmpr4}, the relation between the radius modulus and its dual follows from (\ref{tt1}), was obtained in:
\begin{equation}
\label{trans}
\widetilde{\mathtt{R}}= \frac{\vert \gamma\tau+\alpha\vert \vert q\tau-p\vert ^{2/3}}{T_{\mathtt{M2}}^{2/3}(Im \tau)^{4/3}(2\pi)^{4/3}{\mathtt{R}}} \,.
\end{equation}
'T-duality' defines a nonlinear transformation on the charges of the supermembrane since $\mathcal{T}$ is constructed from them, in distinction with the usual $SL(2,Z)$ action on the moduli which is a linear one. This property will be very relevant for understanding the dual multiplet structure. The KK modes are mapped onto the winding modes and viceversa as expected. This property together with the condition $\mathcal{Z}\widetilde{\mathcal{Z}}=1$ ensure that $(\textrm{T-duality})^2= {I}$. This transformation becomes a symmetry for $\mathcal{Z}=\mathcal{\widetilde{Z}}=1$ which imposes a relation between the tension, the moduli and the KK charges of the wrapped supermembrane,
\begin{equation}
\label{la}
 T^0_{\mathtt{M2}}=\frac{\vert q\tau-p\vert}{{\mathtt{R}}^3 (Im\tau)^2}\, .
\end{equation}
Given the values of the moduli it fixes the allowed tension $T^0_{\mathtt{M2}}$ or on the other way around, for a fixed tension $T^0_{\mathtt{M2}}$, the radius, the Teichm\"uller parameter of the 2-torus, and the KK charges satisfy (\ref{la}). For $\mathcal{Z}=1$ the Hamiltonian and the mass operator of the supermembrane with central charges are invariant under 'T-duality':
\begin{equation}
\begin{aligned}
&{\bf  M}^2 = (T^0_{\mathtt{M2}})^2 {\mathtt{n}}^2 \mathtt{A}^2 + \frac{k^2}{Y^2}+ (T^0_{\mathtt{M2}})^{2/3}H = \\
&\qquad=\frac{{\mathtt{n}}^2}{\widetilde{Y}^2}+ (T^0_{\mathtt{M2}})^2 k^2 \widetilde{\mathtt{A}}^2+ (T^0_{\mathtt{M2}})^{2/3}\widetilde{H} \,,
\end{aligned}
\end{equation}
with $H=\widetilde{H}$. See \cite{gmpr} for further details.
In \cite{gmpr4} the authors showed that there always exists a $\mathcal{T}$ a parabolic matrix transformation of 'T-duality' given for any arbitrary value of the KK  and winding charges. This parabolic transformation depends on the winding and KK momenta of the supermembrane bundle. In this work we extend this analysis to characterize in detail the most general 'T-duality' transformation $\mathcal{T}$, see the appendix C. In principle the 'T-duality' transformations are a subset of the parabolic, elliptic and hyperbolic transformations with equal diagonal terms. We show there that the parabolic one plays a distinguished role since it is the only class able to map any kind of winding and KK charges for a given supermembrane torus bundle with arbitrary monodromy and general central charge into its dual. If we restrict the central charge to $\mathtt{n}=1$, then all the different types of 'T-dual' transformations $\mathcal{T}$, are allowed, i.e. elliptic $\mathcal{T}_{e}$, parabolic $\mathcal{T}_p$ and hyperbolic ones $\mathcal{T}_h$ however, since the supermembrane can wrapped any arbitrary times a 2-torus so there is no justification to restrict to $\mathtt{n}=1$. For $\mathtt{n}\ne 1$ the situation is different: the $\mathcal{T}_{Z_4}$ elliptic case of 'T-duality' transformations fail to map the torus-bundles except for very specific KK and W charges. In distinction, the hyperbolic 'T-duality' matrices $\mathcal{T}_h$, always exist for arbitrary $\mathbf{Q}$, $\mathbf{W}$, and $\mathtt{n}$. The difference with respect to the parabolic case $\mathcal{T}_p$, relies in the fact that for each set of charges there is needed a different hyperbolic realization with different trace. For a general transformation, the parabolic 'T-duality' one $\mathcal{T}_p$ is going to be the one responsible to characterize the different M2-torus bundles dual which at low energies are associated with the different type II gauged supergravities in nine dimensions, for that reason we will denote $\mathcal{T}_p\equiv \mathcal{T}$.
%

%%%%%%%%%%%%%%%%%%%%%%%%%%%%%%%%%%%%%%%%%%%%%%%%%%%%%%%%%%%%%%%%%%%%%%%%%%%%%%%%%%%%%%%%%%%%
%
%
\section{Classification of U-dual supermembrane bundles}
In this section we are going to establish the precise correspondence between the type IIA side of the supermebrane bundle with parabolic, elliptic, hyperbolic and trombone monodromies.  We will focus on the 'T-duality' action analyzed in the preceding section.
\begin{figure*}
\includegraphics[width=0.75\textwidth]{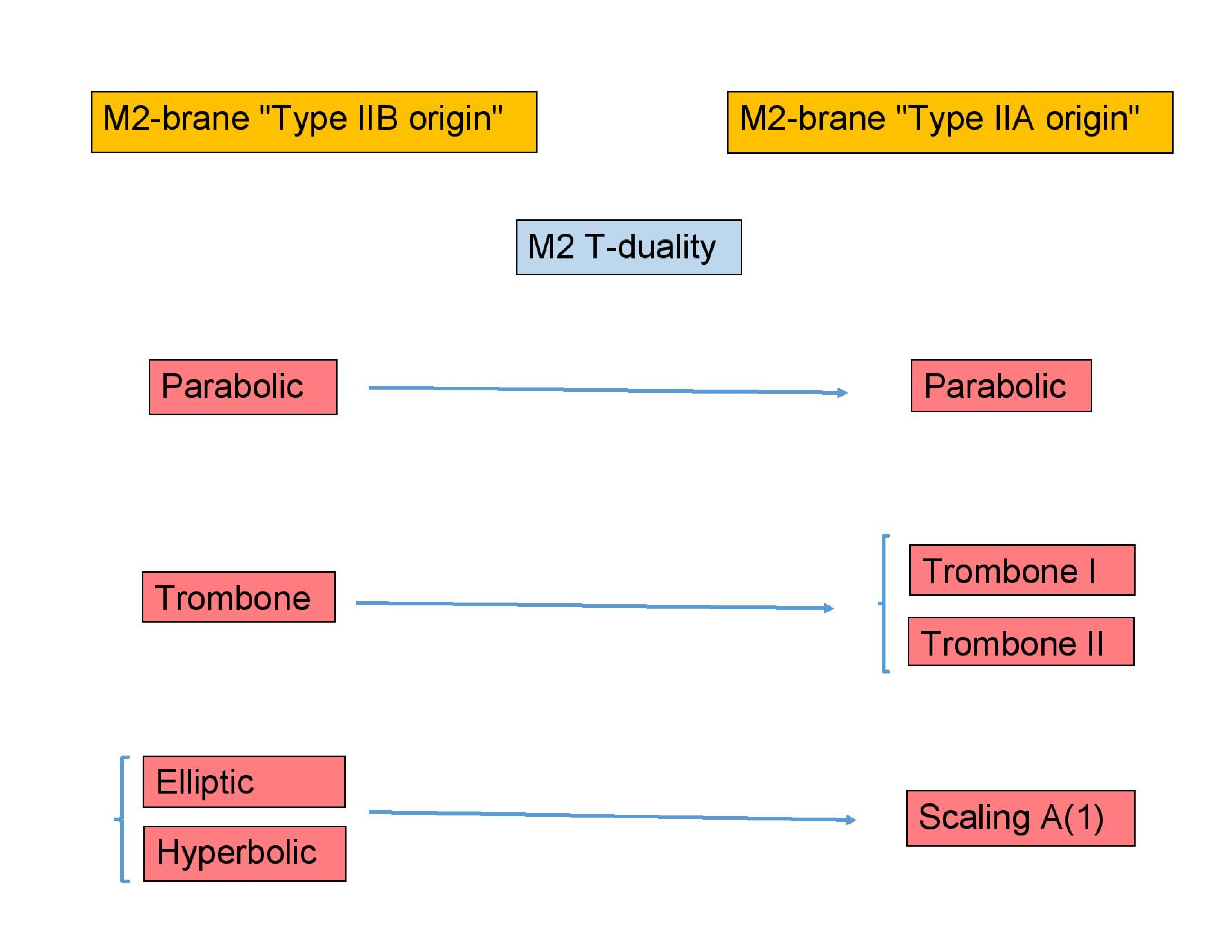}
\caption{\it These are the precise relations between the M2-brane bundle with monodromy in $SL(2,Z)$ inequivalent classes when a M2-brane 'T-duality' is performed. \label{fig:dibujoborrador4}}
\end{figure*}
%
%%%%%%%%%%%%%%%%%%%%%%%%%%%%%%%%%%%%%%%%%%%%%%%%%%%%%%%%%%%%%%
 The M2-brane bundle analysis explains this fact since the 'T-dual' transformation in general does not commute with the monodromy group (except for the parabolic monodromy case) and consequently its associated 'T-dual' coinvariant class of the bundle does not lie in same equivalence class of the original one. This happens even though the monodromy and its dual are in the same conjugation class as originally signalled in \cite{hullgauged}. 

The 'T-dual' transformation maps a given charge $\mathbf{Q}$ in the KK sector into a winding $\mathbf{W}=\mathcal{T}\mathbf{Q}$ and the coinvariant class of $\mathbf{Q}$, to the coinvariant class of $\mathbf{W}$. That is, the coinvariant class
\begin{equation}
C_F\equiv \{\mathbf{Q}+(\mathcal{M}_G-\mathbb{I})\widehat{\mathbf{Q}}\} \,  ,
\end{equation}
is mapped into
\begin{equation}
\widetilde{C}_F=\{\mathbf{W}+(\mathcal{M}^*_G-\mathbb{I})\widehat{\mathbf{W}}\} \,=C_B  ,
\end{equation}
on the winding sector, which may or may not coincide with
\begin{equation}
\mathcal{T} C_F=\{ \mathcal{T} \mathbf{Q}+\mathcal{T}(\mathcal{M}_G-\mathbb{I})\mathcal{T}^{-1}(\mathcal{T}\widehat{\mathbf{Q}})\} \,  ,
\end{equation}
because generically
\begin{equation}
{\mathcal{M}}_G^C:=\mathcal{T} {\mathcal{M}}_G \mathcal{T}^{-1} \ne {\mathcal{M}}_G^*=\Omega \, {\mathcal{M}}_G \,\Omega^{-1} \, .
\end{equation}
If both classes define the same coinvariant class, i.e. $\widetilde{C}_F=\mathcal{T}C_F$, then for any element of the matrix group monodromy ${\mathcal{M}}_{g_1}^*$ and $\widehat{{\mathbf{W}}}$
\be
\label{consistency}
({\mathcal{M}}_G^C-\mathbb{I})\mathcal{T}\widehat{{\mathbf{Q}}}-({\mathcal{M}}_{g_1}^*-\mathbb{I})\widehat{W}=({\mathcal{M}}_{g_2}^*-\mathbb{I})\widehat{\widehat{{\mathbf{W}}}} \, ,
\ee
for some ${\mathcal{M}}_{g_2}^*$, $\widehat{\widehat{{\mathbf{W}}}}$ denotes an arbitrary element in the winding sector. ${\mathcal{M}}_{g_1}^*, {\mathcal{M}}_{g_2}^* \in {\mathcal{M}}_G^*$. In this case, the 'T-duality' map transforms the symplectic torus bundles with a given monodromy group onto themselves. Generically that is not the case, i.e., $\widetilde{C}_F\ne \mathcal{T}C_F,$ so we have to study in detail what occurs for each type of bundle of monodromy: parabolic, elliptic, hyperbolic and trombone.
%
%
%%%%%%%%%%%%%%%%%%%%%%%%%%%%%%%%%%%%%%%%%%%%%%%%%%%%%%%%%%%%%%%%%%%%%%%%%%%%%%%%%%%%%%%%%%%%%%%%%%%%%%%%%%%%%%%%%%%%%%%%%%%%%%%%%
\subsection{U-dual of M2-brane parabolic torus bundles}
The M2-brane torus bundles with parabolic monodromies has two inequivalent nontrivial monodromies classes
\begin{equation}
{\mathcal{M}}_{p} =
  \left( \begin{array}{cc} 1 & \mathtt{p}\\0 & 1\end{array} \right), \quad
{\mathcal{M}}_{Z_2}=\left( \begin{array}{cc} -1 & 0\\
0 & -1\end{array} \right) \,,
\end{equation}
satisfying that $\vert Tr({\mathcal{M}}_p)\vert=2$.
While associated to the first type of monodromy the coinvariants are generically torsionless (the only exception is the $(0,0)$ class associated to torsion one), in the second type of parabolic monodromy all coinvariants have torsion and the KK charges are restricted to be $(0,0),(0,1), (1,0),(1,1)$. A coinvariant with torsion determines a symplectic torus bundle whose local symplectic structure on the fibers is the restriction of the global symplectic one. For the parabolic case the dual monodromy group  is given by ${\mathcal{M}}_p^C= \mathcal{T} {\mathcal{M}}_g \mathcal{T}^{-1}= {\mathcal{M}}_p$ since $\mathcal{T}$ commute with ${\mathcal{M}}_p$. In addition the group ${\mathcal{M}}^*_p= \Omega {\mathcal{M}}_p \Omega^{-1}$ coincides with the group ${\mathcal{M}}_p$ since $\Omega g\Omega^{-1}=g^{-1}$ for any $g\in {\mathcal{M}}_p$. The condition (\ref{consistency}) is then satisfied. The 'T-duality' map acts linearly and it transforms the class of the M2-brane symplectic torus bundles with parabolic monodromy group onto itself.  The same occurs for  ${\mathcal{M}}_{Z_2}$. They are the unique conjugation class of bundles satisfying this property. At low energies we conjecture that they correspond to a parabolic gauged supergravity on the type IIA side. This 9D gauged supergravity is associated to the KK reduction of the 10D Romans supergravity \cite{bergshoeff}.
%
%
%%%%%%%%%%%%%%%%%%%%%%%%%%%%%%%%%%%%%%%%%%%%%%%%%%%%%%%%%%%%%%%%%%%%%%%%%%%%%%%%%%%%%%%%%%%%%
%
%
\subsection{U-duals of M2-brane elliptic and hyperbolic torus bundles}
We will analyze the two cases separately and we will see that under 'T-duality' they generate a unique class of supermembrane dual torus bundles.

The elliptic monodromy group ${\mathcal{M}}_e$ \cite{dabholkar-hull} is finitely generated by the following matrices
\begin{equation}
\begin{aligned}
&{\mathcal{M}}_{Z_3}=\left( \begin{array}{cc} 0 & 1\\ -1 & -1 \end{array} \right)^{\gamma}
\,,
\qquad {\mathcal{M}}_{Z_4}=\left( \begin{array}{cc} 0& 1\\ -1 & 0\end{array}
\right)^{\gamma}\,,\\
&{\mathcal{M}}_{Z_6}=\left( \begin{array}{cc} 1 & 1\\- 1 & 0\end{array}
\right)^{\gamma}\,,
\end{aligned}
\end{equation}
satisfying that $\vert Tr({\mathcal{M}}_e)\vert<2$ with $\gamma\in Z$. Their associated coinvariant of the fiber equivalence classes can be computed explicitly using (\ref{coinvariantclasskk}). For the $Z_3$ case there are three classes
\begin{equation}
C^F_{Z_3}=\{{\mathbf{Q}}_i+({\mathcal{M}}_{Z_3}-\mathbb{I})\widehat{{\mathbf{Q}}}\},\quad i=1,2,3 \,,
\end{equation}
with
\begin{equation}
{\mathbf{Q}}_1=\left( \begin{array}{c} 0 \\ 0 \end{array}\right)\,, \quad {\mathbf{Q}}_2= \left( \begin{array}{c} 0\\ 1\end{array}\right)\,, \quad
{\mathbf{Q}}_3=\left( \begin{array}{c} 0\\ 2\end{array}\right)\,.
\end{equation}

It can be shown that (\ref{consistency}) cannot be satisfied. In fact (\ref{consistency}) means that for any $\widehat{g}\in {\mathcal{M}}_{Z_3}^C$ and $\widehat{{\mathbf{Q}}}$, there exists $g\in {\mathcal{M}}^*_{Z_3}$ and $\widehat{{\mathbf{W}}}$ such that
\begin{equation}
\label{coin}
(\widehat{\mathcal{M}}_g-\mathbb{I})\, \mathcal{T} \,\widehat{{\mathbf{Q}}}=(g-\mathbb{I})\widehat{{\mathbf{W}}} \,.
\end{equation}
One can show that there does not exist $\widehat{{\mathbf{W}}}$ satisfying the equality (\ref{coin}) for suitable $\widehat{g}\in {\mathcal{M}}_{Z_3}^C$ and $\widehat{{\mathbf{Q}}}$.
In the same way it can be shown that (\ref{consistency}) is not satisfied for the monodromy groups ${\mathcal{M}}_{Z_4}, {\mathcal{M}}^*_{Z_4}$.

In distinction (\ref{consistency}) is satisfied for ${\mathcal{M}}_{Z_6}$ and ${\mathcal{M}}_{Z_6}^*$. This means that given any element $\widehat{g}\in {\mathcal{M}}_{Z_6}^C$ and any $\widehat{\mathbf{Q}}$ there exists $g\in {\mathcal{M}}^*_{Z_6}$ and $\widetilde{{\mathbf{W}}}$ satisfying (\ref{consistency}) and viceversa. This follows because one element of the group is $g_{Z_6}=\left( \begin{array}{cc} 1 & 1\\- 1 & 0\end{array} \right) \,,$ hence $(g^*_{Z_6}-\mathbb{I})$ has determinant equal to 1, then there always exists $\widetilde{W}$ satisfying (\ref{consistency}):
\begin{equation}
(g^*_{Z_6}-\mathbb{I})^{-1}(\widetilde{g}^*_{Z_6}-\mathbb{I})\,\mathcal{T}\, \widehat{{\mathbf{Q}}}=\widetilde{{\mathbf{W}}}.
\end{equation}
The other way round follows in the same way, given any $g\in {\mathcal{M}}^*_{Z_6}$ and $\widetilde{{\mathbf{W}}}$ there always exists $\widehat{{\mathbf{Q}}}$ satisfying (\ref{coin}). In fact
\begin{equation}
det (\mathcal{T}g_{Z_6}\mathcal{T}^{-1}-\mathbb{I})=1 \,,
\end{equation}
and we proceed as before. The coinvariant abelian group associated to the monodromy ${\mathcal{M}}_{Z_6}$ has only one element in the KK sector given by the class
\begin{equation}
C^F_{Z_6}=\{{\mathbf{Q}}_0+({\mathcal{M}}_{Z_6}-\mathbb{I})\widehat{{\mathbf{Q}}}\} \,,
\end{equation}
where ${\mathbf{Q}}_0$ is any particular charge and respectively the only one element in the winding sector.

%%%%%%%%%%%%%%%%%%%%%%%%%%%%%%%%%%%%%%%%%%%%%%%%%%%%%%%%%%%%%%%%%%%%%%%%%%%%%%%%%%%%%%%
%
Let us now analyze the case of hyperbolic monodromy. There are infinite abelian monodromy groups of hyperbolic matrices constructed in terms of
\begin{equation}
{\mathcal{M}}_{h}=\left( \begin{array}{cc} a & b\\ c & d \end{array} \right)^{\gamma} \,,
\end{equation}
such that $\vert Tr({\mathcal{M}}_h)\vert>2$ with ${\mathcal{M}}_h\in SL(2,Z)$ \cite{dabholkar-hull}. It can also be explicitly shown that generically the coinvariant structure of the dual bundle is not equivalent to the original one.

Let us now compare the dual monodromies of elliptic and hyperbolic 2-torus bundles  with respect to their conjugate classes. We consider the generic monodromy case $g\in {\mathcal{M}}_G$ with $g=\left( \begin{array}{cc} a & b\\ c & d \end{array} \right)\in SL(2,Z)\,,$ then
\begin{equation}
\mathcal{T} \, g \, \mathcal{T}^{-1}=\left( \begin{array}{cc} a+tc & \quad-t(a+tc)+b+td\\ c & -ct+d \end{array} \right)\,.
\end{equation}
By defining with  $u=a-d$, we can always express
\begin{equation}
\label{alg} 
(\mathcal{T} g \mathcal{T}^{-1}-\mathbb{I})-((g^{*})^{-1}-\mathbb{I})= (u+tc)[B-tA] \,,
\end{equation}
with $A=\begin{pmatrix}0&1\\0&0\end{pmatrix}$ and $ B=\begin{pmatrix}-1&0\\0&1\end{pmatrix}.$ Notice that $A,B$, satisfy the algebra:
\begin{equation}
[A,B]=2A \,,
\end{equation}
which corresponds to a non abelian group $A(1)$ associated to the collinear transformations in one dimension (translation and scaling). The same  algebra was already identified at the level of nine dimensional type II gauged supergravity in the 'type IIA sector'\cite{bergshoeff}.

%%%%%%%%%%%%%%%%%%%%%%%%%%%%%%%%%%%%%%%%%%%%%%%%%%%%%%%%%%%%%%%%%%%%%%%%%%%%%%%%%%%%%%%%%
%
Notice that for the parabolic case $(a=1, b=\mathtt{p}, c=0, d=1)$ both coefficients in (\ref{alg}) vanish, and the coinvariants and their duals lie in the same equivalence class.

Clearly the elliptic and hyperbolic coinvariant classes are mapped under 'T-duality' into an inequivalent coinvariant equivalence class. 'T-duality' acts nonlinearly on the supermembrane dual bundle and  this dual realization is associated to a nonabelian algebra of the $A(1)$ group. We conjecture that the A(1) M2-brane 2-torus bundle at low energies is described  by the type IIA gauged supergravities corresponding to a nonabelian $A(1)$.
%%%%%%%%%%%%%%%%%%%%%%%%%%%%%%%%%%%%%%%%%%%%%%%%%%%%%%%%%%%%%%%%%%%%%%%%%%%%%%%%%%%%%%%%
%

\subsection{U-dual of M2-brane trombone torus bundles}
Trombone symmetry produces supergravities that do not have Lagrangian but are uniquely defined through the equations of motion. The reason is that the trombone symmetry is not a symmetry of the action since it scales the Langrangian but it is a symmetry of the equations of motion. At quantum level however there exists a well defined action since it is possible to define an invariant hamiltonian. In type II supergravity in 9D the global symmetries are $GL(2,R)=SL(2,R)\times R$, the breaking of the group into its arithmetic subgroup $GL(2,Z)$ with determinant $\pm 1$ is not able to capture the effect of the scaling. Hence in order to obtain the scaling symmetries an alternative procedure is needed.  This question was in fact solved many years ago by the authors \cite{cremmer}, by means of a nonlinear realization of the group $SL(2,Z)$ that they called \textit{active $SL(2,Z)$}.  In \cite{gmpr} the authors used this realization to obtain a bundle description of the supermembrane with gauged \textit{trombone} symmetry. In the following, in order to be self-contained we first summarize those results i.e. the realization of the trombone symmetry and their associated torus bundles, previous to perform the characterization of their duals.

Let us consider a nonlinear representation of the group $SL(2,Z)$ in terms of the $2\times 2$ matrices ${\mathbb{H}}_{ij}$ as in \cite{gmpr}. Given two different charges  $\mathbf{Q}_i$, $\mathbf{Q}_j$ labelled with two different indices $i,j$ and given ${\mathbb{H}}_{ij}$ the $SL(2,Z)$ active transformation, it acts on the charges as follows:
\begin{equation}
\label{trombone}
{\mathbb{H}}_{ij}\mathbf{Q}_i=\mathbf{Q}_j, \quad \textrm{and}\quad \frac{{\mathbb{H}}_{ij}}{h_{ij}}\left( \begin{array}{c} \tau\\ 1\end{array} \right)=\left( \begin{array}{c} \tau\\ 1\end{array} \right)\, ,
\end{equation}
the solution for (\ref{trombone}) is given by
\begin{equation}
{\mathbb{H}}_{ji}=
\left( \begin{array}{cc}
-\frac{p_j}{q_j}U+\frac{q_i}{q_j}C \quad & \frac{p_i}{q_i}+\frac{p_ip_j}{q_iq_j}U-\frac{p_i}{q_j}C
\\ -U & \frac{q_j}{q_i}+\frac{p_i}{q_i}U\end{array} \right)\,
\end{equation}
where
{\small
\begin{equation}
h_{ji}=\frac{p_j-q_j\overline{\tau}}{p_i-q_i\overline{\tau}};\quad U=\frac{p_jq_i-p_iq_j}{\vert p_i-q_i\overline{\tau}\vert^2};\quad C=\frac{\vert p_j-q_j\tau\vert^2}{\vert p_i-q_i\tau\vert^2} \,.
\end{equation}}
For each monodromy group there exists a unique non linear realization of it ${\mathbb{H}}_G$. There are three nonlinear realizations associated to the elliptic, parabolic and hyperbolic monodromy classes, however they cannot be distinguish among them, so on the type IIB side we only obtain a single M2-brane trombone torus bundle class. This is in agreement with the fact that at low energies on the type IIB side there is a unique 9D trombone supergravity.

This transformation generates the complete lattice of charges for a given vacuum, (that is the asymptotic value of the scalar moduli).  ${\mathbb{H}}_{ji} \in GL(2,R)$ is the nonlinear representation of $\mathcal{M}_{ji}\mathbf{Q}_i=\mathbf{Q}_j$ with $\mathcal{M}_{ji}\in SL(2,Z)$ which acts linearly on the charges but non-linearly on the moduli. The Hamiltonian $H$ is invariant, since
\begin{equation}
\tau\to \tau,\quad \mathbf{W}\to \mathbf{W},\quad \mathtt{R}\to \mathtt{R}\ \, .
\end{equation}
While the mass operator changes since the KK contribution changes, $KK\to KK'$ according to (\ref{trombone}). See \cite{gmpr} for details.\newline
The structure of the supermembrane trombone bundle differs from those in which the monodromy is linearly realized since in the former the moduli is unaltered by the trombone monodromy. Consequently on the 'type IIB side' each bundle only contains a single pair of winding charges $\mathbf{W}=\left(\begin{array}{cc} l\\ m \end{array}\right)$ instead of an orbit, and therefore there is no monodromy associated to the base manifold. The only monodromy of the bundle is associated to the fiber $\mathcal{M}_G$, it is non linearly realized in terms of ${\mathbb{H}}_G$. The coinvariant class of charges of the fiber is defined as
\begin{equation}
C_F^{tromb}=\{\mathbf{Q}+({\mathbb{H}}_G-\mathbb{I})\widehat{\mathbf{Q}}\} \,,
\end{equation}
and the structure of the bundle in terms of the coinvariant classes is $(C_F^{tromb}, \mathbf{W})$.
The 'T-duality' transformation for the trombone torus bundle maps the coinvariant class of the fiber of the original bundle into the coinvariant class of the base, and the original winding single charge of the base into a single dual KK charge of the dual fiber. As a result, the dual trombone bundle $(\widetilde{C}_B^{tromb},\mathbf{Q})$, corresponds to a bundle that has a trivial monodromy on the fiber but a nontrivial monodromy in the base manifold with  coinvariant classes given by
\begin{equation}
\widetilde{C}_B^{tromb}=C_F^{tromb} \, ,
\end{equation}
$\widetilde{\tau}$ y $\widetilde{\mathtt{R}}$  the 'T-dual' moduli parameters transform as in (\ref{tt1}) and (\ref{trans}), they are defined by the covariant class. They do not depend on any particular element of the orbit.  
The geometric structure can be interpreted as a compatible set of fiber bundles with characteristic classes defined by the coinvariant class of winding matrices.
Under the 'T-duality' transformation the hamiltonian which is invariant on the orbits of ${\mathbb{H}}_G$ is transformed to a hamiltonian invariant under the orbits of ${\mathbb{H}}^{\ast}_G$ and the bundle change as 
$(C_F^{tromb}, \mathbf{W})\stackrel{\mathbb{T}}{\rightarrow} (\widetilde{C}_B^{tromb},\widetilde{\mathbf{Q}}) $.

We notice that in the expression of the hamiltonian of the trombone torus bundle with coinvariant $C_F^{trombone}$ there is not a $\Theta$ matrix on the expression of the covariant derivative, since $\mathbf{W}, \tau, \mathtt{R}$ remain fixed under the action of the nonlinear transformation ${\mathbb{H}}_G$. At low energies they correspond to gauged theories in nine dimensions obtain of the massive deformation of type IIA supergravity in 10D. However there exists a $\Theta$ matrix in the covariant derivatives which compensates the transformation of the winding matrix on the dual torus bundle ('type IIB' side).
%
%
%
%
%%%%%%%%%%%%%%%%%%%%%%%%%%%%%%%%%%%%%%%%%%%%%%%%%%%%%%%%%%%%%%%%%%%%%%%%%%%%%%%%%%%%%%%
%
%
\subsubsection{Two inequivalent duals of the M2-brane trombone torus bundle}
The gauging is obtained by means of the nonlinear representation ${\mathbb{H}}^G$ with $\mathcal{M}_G\subset \mathcal{M}_{ji}$ associated to a particular monodromy equivalence class. To compute it we particularize the monodromy matrix $\mathcal{M}_G \mathbf{Q}_i=\mathbf{Q}_j$ and we substitute\newline ${\mathbb{H}}_{ji}(\tau,q_i,p_i,q_j,p_j)$ and $\mathbf{Q}_i=\left( \begin{array}{c}p_i\\q_i\end{array} \right)\,$  and respectively $\mathbf{Q}_j$.
The linear representation $\mathcal{M}_{ji}$ contains the three inequivalent conjugation classes (elliptic, parabolic and hyperbolic) however the nonlinear representation ${\mathbb{H}}_{ji}$, does not have any particular property which may distinguish the elliptic to the hyperbolic cases. Under 'T-duality' in distinction with the previous case, there are two inequivalent types of dual principal bundles, one with  the parabolic monodromy non linearly realized in the base  and a second one associated to the nonlinear realization elliptic and hyperbolic monodromy dual in the base manifold. In the case of parabolic monodromy the duality matrix $\mathcal{T}$ commute with ${\mathbb{H}}_{ij}$. In the other two cases the 'T-duality' does not commute with ${\mathbb{H}}_{ij}$ but since they cannot be distinguish they form a second inequivalent class of trombone torus bundles on the 'type IIA side'. Nicely at low energies they are in correspondence with the two trombone gauged supergravities of the type IIA sector.
%
%%%%%%%%%%%%%%%%%%%%%%%%%%%%%%%%%%%%%%%%%%%%%%%%%%% %%%%%%%%%%%%%%%%%%%%%%%%%%%%%%%%%%%%%
%
%

\section{Discussion and Conclusions}

We find eight independent inequivalent classes of supermembrane torus bundles with monodromy linearly and non linearly realized in $SL(2,Z)$. Four of them had been previously found, but those associated to the type IIA side are new. The M2-brane torus bundle is characterized by the monodromy, the coinvariants of the base manifold $C_B$ and that of the fiber $C_F$.  Their Hamiltonians and mass operators for the nine different cases (eight inequivalent class of bundles with monodromy and one principal) are all invariant under the duality action that we have defined in this work irrespective on the monodromy group. However the structure of the bundle, for arbitrary M2-brane torus bundles with monodromy,  is not necessarily preserved. Only for the cases in which U-duality acts linearly (for example for the monodromy is parabolic or trivial ) the bundle structure is  preserved. We observe that the monodromy group of the M2-brane torus bundle coincides with the gauging groups of type II supergravity.  Hence we relate each of these inequivalent classes of M2-brane 2-torus bundles  -with  and without monodromy- with each of the eight type II gauged supergravities in 9D and the maximal one respectively. Recently it has been  shown that the  supermembrane theory with central charge condition (described by 2-torus bundles with monodromy) is equivalent to a supermembrane on a constant three-form background toroidally compactified in the presence of \textit{fluxes}  \cite{gmlhlpr}. Since flux compactifications are related to gauged supergravities this is another way to evidence the relation between the M2-brane torus bundle with monodromy and its description at low energies in terms of gauged supergravities. The non-preservation of the coinvariant equivalence classes under U-duality we believe that is the underlying reason for the different structure between the type IIB and type IIA gauged supergravities in nine dimensions from the M-theory viewpoint. 
We have proved that the U-dual of the parabolic coinvariants remains in the same class of coinvariants, moreover, they seem to be the only ones that preserve the coinvariant class. Consequently the type IIB parabolic M2-brane bundle is mapped through duality into the parabolic type IIA M2-brane bundle. At low energies it becomes  natural to relate it to the type IIA parabolic supergravity, which corresponds to the KK reduction of Romans supergravity in 10D to 9D.

The U-duals of the supermembrane with monodromies elliptic and hyperbolic do not preserve the coinvariant class, even though the dual monodromy is conjugated to the original one. U-duality acts non linearly on these M2-brane bundles and it is responsible for changing the bundle class. The algebra of the monodromy groups elliptic and hyperbolic with the U-duality transformation form a non-abelian algebra A(1) that acts like an scaling and a translation. The dual M2-brane bundle with these two inequivalent classes of monodromies  corresponds to a single class of dual M2-brane bundles that we call $A(1)$.  At low energies we consider that it could be related it to the type IIA $A(1)$ gauged supergravity in 9D. 

Trombone symmetry of the supermembrane torus bundle is a non linear realization of the $SL(2,Z)$ group. It acts linearly on the charges but non-linearly on the moduli and the bundle description once it is gauged is completely different to the previous cases considered. The M2-brane bundle with gauged trombone monodromy are classified according to the monodromy, but they contain only one class of coinvariants either of the fiber ('type IIB side') or of the base ('type IIA side'). Under U-duality the nonlinear realization of the three inequivalent classes of M2-brane torus bundles (elliptic, hyperbolic or parabolic subgroups) they form a unique M2-brane trombone bundle on the 'type IIB' side. Under U-duality it maps into two inequivalent classes of dual M2-brane trombone bundles on the 'type IIA' side: one associated to the gauging of the non linear realization of the parabolic monodromy and a second one associated to the gauging of the non linear realization of the bundles with elliptic and hyperbolic monodromy. Nicely these two inequivalent classes of T-bundles can naturally be associated on the 10D 'type IIA side' to the  KK reduction of the massive Howe, Lambert and West supergravity and to the SS reduction of type IIA maximal supergravity. 
%%%%%%%%%%%%%%%%%%%%%%%%%%%%%%%%%%%%%%%%%%%%%%%%%%%%%%%%%%%%%%%%%%%%%%%%%%%%%%%%%%%%%%%%%%%%%%%%%%%%%%%%
%
%
%
%

%
\section{Acknowledgements}
 JM Pena is grateful to the Mainz Institute for Theoretical Physics (MITP) for its hospitality and its partial support during part of the realization of this work. MPGM is grateful to I. Cavero-Pelaez for helpful comments on the manuscript. MPGM has been supported by Mecesup ANT1398 and ANT1555, Universidad de Antofagasta (Chile). AR and MPGM are partially supported by Projects Fondecyt 1161192 (Chile). MPGM and AR participated as external nodes of the EU-COST Action MP1210 `The String Theory Universe' during the realization of this work.

\appendix
%%%%%%%%%%%%%%%%%%%%%%%%%%%%%%%%%%%%%%%%%%%%%%%%%%%%%%%%%%%%%%%%%%%%%%%%%%%%%%%%%%%%%%%%%%%%%%%%%%%%%%%%%%%%%5
%%%%%%%%%%%%%%%%%%%%%%%%%%%%%%%%%%%%%%%%
\section{Coinvariants}
Let us define the map associated to the coinvariant classes on the KK sector to the coinvariant classes on the winding sector. We take one element ${\mathbf{Q}}$ in a coinvariant class of the KK sector and map it to an element ${\mathbf{W}}$ of a coinvariant class of the winding sector. There exists an element of $SL(2,Z)$ with equal diagonal terms $\mathcal{T}$ which maps ${\mathbf{Q}}$ to ${\mathbf{W}}$ :
\begin{equation}
\mathbf{W} =\mathcal{T} \mathbf{Q}.
\end{equation}
In the case, as we are considering here, where the abelian group $G$ has only one generator $J$:
\begin{equation}
g\in {\mathcal{M}}_G,\quad g= J^a \, ,
\end{equation}
where $a$ is an integer, then, $g-\mathbb{I}= J^a-\mathbb{I}$. We can distinguish three different cases:
\begin{equation}
\bea
&\textrm{if} \quad a=0\,,\qquad g-\mathbb{I}=0 \,, \\
&\textrm{if}\quad a>0 \,, \qquad J^a-\mathbb{I}=(J-\mathbb{I})(J^{a-1}+J^{a-2}+\dots+\mathbb{I})\,, \\
&\textrm{if} \quad a<0 \,, \qquad J^a-\mathbb{I}=(\mathbb{I}- J^{-a})J^a \,.
\eea
\end{equation}
Hence if $a=0$ we have one element ${\mathbf{Q}}$.
If $a>0$ we have
\begin{equation}
C_F=\{{\mathbf{Q}}+(J-1) {\hat{\hat{{\mathbf{Q}}}}}\} ,
\end{equation}
where $\hat{\hat{\mathbf{Q}}}=[J^{a-1}+J^{a-2}+\dots+\mathbb{I}]\, \hat{{\mathbf{Q}}}.$

If $a=1$ we get $\hat{\hat{\mathbf{Q}}}=\hat{\mathbf{Q}}$ and if $a\ge 2$, $\hat{\hat{\mathbf{Q}}}$ belongs to a subset of the whole $\hat{\mathbf{Q}}$ sector. We then conclude that the coinvariant class can be expressed as
\begin{equation}
\label{elements}
C_F=\{{\mathbf{Q}}+(J-1) {\hat{\mathbf{Q}}}\} .
\end{equation}
The same result holds for $a<0$. Consequently without loose of generality we can take the elements of the coinvariant class as (\ref{elements}). Notice that the case $a=0$ is also contained in the class since it corresponds to consider $\hat{{\mathbf{Q}}}=\begin{pmatrix}
                            0 \\
                           0
                          \end{pmatrix}.$
%\newline
%Using the same arguments we get $C_B=\{{\mathbf{W}}+\Omega(J-1)\Omega^{-1}\hat{{\mathbf{W}}}\}.$
%%%%%%%%%%%%%%%%%%%%%%%%%%%%%%%%%%%%%%%%%%%%%%%%%%%%%%%%%%%%%
\section{General 'T-duality' transformation}
\label{appC}
In this appendix we obtain the most general transformation of 'T-duality' for the supermembrane T-bundles.  Given a particular supermembrane wrapping the 2-torus target space it has associated a matrix of windings $\mathbb{W}$ with determinant $det(\mathbb{W})=\mathtt{n}\ne 0$, such that applying an $\mathbb{S}\in SL(2,Z)$ it admits a triangular description,
\begin{equation}
\mathbb{W}\, \mathbb{S}=\begin{pmatrix}\mathtt{n} & e\\ 0 & 1\end{pmatrix} \,.
\end{equation}
Multiplying on the left hand side by a parabolic matrix $\mathbb{R}$ the winding can be expressed in terms of its canonical form $\mathbb{R}\,\mathbb{W} \, \mathbb{S}=\begin{pmatrix}\mathtt{n} & 0\\ 0 &1\end{pmatrix}$.
On the other hand, in \cite{gmpr5} the authors showed that always exists a map 'T-duality' maps
\begin{equation}
\mathbb{W}\mathbb{S}=\mathcal{T} \mathbb{Q}\,,
\end{equation}
where $\mathbb{Q}$ is the matrix of charge of determinant $\mathtt{n}$ whose first row is $\mathbb{Q}$. $\mathcal{T}$ for sake of brevity denotes the parabolic 'T-duality' transformation $\mathcal{T}_p$. Then, for arbitrary $\mathbb{W},\mathbb{Q}$ charges we want to obtain the more general $\hat{S}, \mathcal{T}$ such that they satisfy $\mathbb{W}\hat{\mathbb{S}}=\mathcal{T}\mathbb{Q}.$
Consequently,
\begin{equation}
\mathbb{W}=\mathcal{T}\mathcal{T}^{-1}\mathbb{W}\mathbb{S}\hat{\mathbb{S}}^{-1}.
\end{equation}
We want to find the more general $u=\mathcal{T}\mathcal{T}^{-1}$ and $v=\mathbb{S}\hat{\mathbb{S}}^{-1}$ such that the above relation is satisfied. This condition is verified for
\begin{equation}
\mathcal{T}=A B A^{-1}(\mathcal{T})=\begin{pmatrix} d-e c& \quad dt-b\mathtt{n}+e(a-d-ct-ce)\\
-c& -c(t-e)+a\end{pmatrix} \,,
\end{equation}
with $A= \begin{pmatrix}1& e\\ 0&1\end{pmatrix}$ and $B=\begin{pmatrix}d& -b\mathtt{n}\\ -c&a\end{pmatrix}$ with $A,B\in SL(2,Z)$.

One can distinguish three different cases according to the value of the trace
\begin{equation}
Trace(\mathcal{T})=d-ct+a \,,
\end{equation}
as elliptic, parabolic and hyperbolic. For the case when the central charge is restricted to be $\mathtt{n}=1$ all of the three subclasses of 'T-duality' matrices are allowed, however this is not the more general case since for the wrapped supermembrane on a torus arbitrary $\mathtt{n}$ is allowed.
\begin{itemize}
\item The parabolic case  $\mathcal{T}_p=
\left( \begin{array}{cc}
1& t\\ 0&1\end{array} \right)$
has already being discussed \cite{gmpr4}, it corresponds to have $a=1, d=1$ with $c=0$ and it is satisfied for any configuration of charges, windings and central charge, so in the following we will analyze the two other possibilities.
\item The elliptic case corresponds to have $c=\pm 1$ and $b\mathtt{n}=\pm 1-e(e-t)$ since it must verify that the two trace entries are equal. For arbitrary central charge $\mathtt{n}\ne 1$ the above equation is not satisfied apart for solutions associated to particular values of charges and windings.
\item For the hyperbolic case we will analyze separately three different cases restricted to assume that $\mathtt{n}$ is prime:
\begin{itemize}
\item In first place let consider $(2e-t)$ and $\mathtt{n}$ relatively prime integers different from zero, then choose $a,b$ such that the following relation is satisfied
\begin{equation}
(2e-t)-b \quad\textrm{then}\quad \mathtt{n}= 1,\quad c=1-a^2 \,.
\end{equation}
In this case the trace is equal to $2(1-a^2)(e-t)+a$, then if $a,b$ are solution then $a+\lambda \mathtt{n}, b+\lambda(2e-t)$ is also a solution. We always can find a solution of reversed sign to $(e-t)$ then $(1-a^2)(e-t)$ and $a$ have the same sign and consequently we only need to choose a $\lambda$ large enough to guarantee that
\begin{equation}
\vert a+\lambda \mathtt{n}\vert >1 \,,
\end{equation}
and it corresponds to an hyperbolic solution. It always exists a hyperbolic solution for this first case.
\item If $(2e-t)$ and $\mathtt{n}$ different from zero are NOT relatively prime. Then, $(2e-t)=m\mathtt{n}$ with $m\ne 0$. Substituting in the relation $ad-bc\mathtt{n}=1$ one obtains $a^2+c(am-b)\mathtt{n}=1.$ There is a solution $b=am$ with $a=1$ and arbitrary $c$. The trace is
\begin{equation}
Tr(\mathcal{T})= 2(e-t)c+a\ge 2.
\end{equation}
One can choose $c=e-t$ and then the trace corresponds to a hyperbolic matrix. The case $e=t$ (which it would not corresponds to an hyperbolic case) cannot happen since in the case we consider $2e-t=m\mathtt{n}$ then $e=m\mathtt{n}$ but $e<\mathtt{n}$ always, as shown in \cite{sl2z}. Consequently it always exists a hyperbolic matrix for this case.
\item The last case to consider occurs when $2e-k=0$ then $a=\mathtt{n}+1, b=\epsilon=sign(e-k), c = \epsilon(\mathtt{n}+2)$ with trace
{\small
\begin{equation}
Tr(\mathcal{T})=2[\epsilon(\mathtt{n}+29(e-k))+a]>2(\mathtt{n}+1)>2 \,.
\end{equation}}
Consequently it always exists a hyperbolic matrix for this case.
\end{itemize}
\end{itemize}

In conclusion there always exists a parabolic transformation  $\mathcal{T}$, and an hyperbolic transformation $\mathcal{T}_h$ inside the set of 'T-dualities' transformations allowed for any value of the central charge $\mathtt{n}$. Each hyperbolic transformation has a different trace for any different set of charges.

\end{document}